\documentclass[5p,authoryear]{elsarticle}

\usepackage{graphics}
\usepackage{graphicx}
\usepackage{epsfig}
\usepackage{rotating}
\usepackage{amssymb}

\usepackage{lineno}

\journal{Journal of Astrophysics and Astronomy, Springer}

\begin{document}

\begin{frontmatter}



\title{LINE SHAPE VARIABILITY IN A SAMPLE OF AGN WITH BROAD LINES}


\author[doa]{D.Ili\'c}
\ead{dilic@math.rs}
\address[doa]{Department of Astronomy, Faculty of Mathematics, University of Belgrade, Studentski trg 16, 11000 Belgrade,
Serbia}
\author[aob,doa]{L.\v C. Popovi\'c}
\address[aob]{Astronomical Observatory, Volgina 7, 11060 Belgrade 74,
Serbia}
\author[sao]{A. I. Shapovalova}
\address[sao]{Special Astrophysical Observatory of the Russian Academy of Science, Nizhnij Arkhyz, Karachaevo-Cherkesia 369167, Russia}
\author[sao]{A. N. Burenkov}
\author[ina]{V. H. Chavushyan}
\address[ina]{Instituto Nacional de Astrof\'{i}sica, \'{O}ptica y 
Electr\'{o}nica, Apartado Postal 51, CP 72000, Puebla, Pue. M\'{e}xico}
\author[doa]{A.Kova\v cevi\'c}

\begin{abstract}

The spectral variability of active galactic nuclei (AGN) is one of their key features that enables us
to study in more details the structure of AGN emitting regions. Especially, the broad line profiles,
that vary both in flux and shape, give us invaluable information about the kinematics and geometry of
the broad line region (BLR) where these lines are originating from.

We give here a { comparative} review of the line shape variability in a sample of five type 1 AGN, those with broad emission lines in their spectra, { of the data obtained from the} international long-term optical monitoring campaign coordinated by the Special Astrophysical Observatory of the Russian Academy of Science. 
The main aim of this campaign is to study the physics and kinematics of the BLR on a uniform data set, focusing on the problems of the photoionization heating of the BLR and its geometry, { where in this paper we give for a first time a comparative analysis of the variabilty of five type 1 AGN, discussing their complex BLR physics and geometry in the framework of the estimates of the supermassive black hole mass in AGN.}

\end{abstract}

\begin{keyword}
galaxies:active-galaxies  quasar:individual (Arp 102B, 3C 
390.3, NGC 5548, NGC 4151, Ark 564) line:profiles



\end{keyword}

\end{frontmatter}


\begin{table*}[t!]
\caption{Spectral characteristics and the variability of the observed objects: object name and the monitoring period,
redshift, AGN type, line profile characteristics and the FWHM of the mean spectrum, photometric BLR radius $R_{\rm BLR}=c\tau_{\rm BLR}$, maximal and minimal flux ratio $F_{\rm max} / F_{\rm min}$, the variability parameter $F_{\rm var}$, mean continuum luminosity at 5100\AA, and the main references. If available
values for both H$\beta$ and H$\alpha$ line are listed.}
\label{tab1}
\vskip 2mm
\centering
\resizebox{18cm}{!}{%
\begin{tabular}{ccccccccc}
\hline
 object  &  z &  AGN   & Line Profile Shape& c$\tau_{\rm BLR}$  [ld] &  $F_{\rm max} / F_{\rm min}$ & F$_{\rm var}$& $\lambda$L$_\lambda$(5100)&  Ref \\
(period [years]) &    &  type  & FWHM  [km/s]   & H${\beta}$/H${\alpha}$ & H${\beta}$/H${\alpha}$  & H${\beta}$/H${\alpha}$ & [$10^{44}$ erg/s] &   \\
\hline
\hline          
 NGC 5548    & 0.0172 &  Sy 1.0-1.8  &  strong shoulders  & 49$^{+19}_{-8}$/27$^{+14}_{-6}$ & 4.9$\pm$0.3/3.5$\pm$0.3 & 0.33/0.30& 0.40$\pm$0.12 & \cite{Sh04} \\
 (1996-2002) &        &                &   6300  &   &    &  &  & \cite{Kov14} \\
\hline
  NGC 4151  & 0.0033 & Sy 1.5-1.8  & absorption,bumps    & 5$^{+28}_{-5}$/6$^{+27}_{-6}$   & 4.8$\pm$0.2/3.1$\pm$0.3  & 0.42/0.31  & 0.05$\pm$0.03 & \cite{Sh08}  \\
(1996-2006) &        &               &   6110/4650    &  &  &  & \\
\hline
  3C390.3  & 0.0561 &  Radio-loud    & double-peaked & 96$^{+28}_{-47}$/120$^{+18}_{-18}$ & 4.7$\pm$0.4/3.4$\pm$0.6 & 0.38/0.35 & 0.90$\pm$0.42  & \cite{Sh10a} \\
(1995-2007) &      &                 &   11900/11000   &   & &  &   \\
\hline
Ark 564       & 0.0247 & NLSy1    &  strong FeII  & 4$^{+27}_{-4}$/5$^{+6}_{-14}$  &1.6$\pm$0.1/1.5$\pm$0.1 & 0.07/0.08 &  0.36$\pm$0.04  &  \cite{Sh12}\\
(1999-2010)   &        &             &  960/800    &    &  &&  & \\
\hline
Arp 102B    & 0.0242   &LINER    &  double-peaked & 15$^{+20}_{-15}$/21$^{+14}_{-14}$  &3.0$\pm$0.2/2.4$\pm$0.2& 0.21/0.20  &0.11$\pm$0.01   & \cite{Sh13}\\
(1987-2013) &        &              & 15900/14300   &    &   &  &       &   \\
\hline
\end{tabular}
}
\end{table*}

\section{INTRODUCTION}

In spite of decades of intensive investigations, the broad line region (BLR) of active galactic nuclei (AGN) 
is yet not fully understood. The direct detection
of the BLR remains a challenge for modern instruments, since the angular size of 
the BLR is less than $0.001\,{\rm arcsec}$ even for the closest AGNs. The only information
from the BLR comes in the form of the broad emission lines (BEL), that are a very prominent feature
in the spectra of the so-called type 1 AGN \citep{OF06}. 
Nevertheless, the investigations of the BEL's properties (flux and shape), 
and especially of their variability, indicates that the BLR is linked with the accretion process 
onto the supermassive black hole in the center of an AGN. First of all, the BLR gas is photoionized by the
continuum radiation from the accretion disk, and secondly, { at least part of the BLR follows its geometry, 
i.e. even though the BLR geometry is not ubiquitously described it is believed that one part of the BLR is 
a part of the accretion disk or has a disk-like geometry \citep{Po04}}.

An important method to indirectly map the geometry and kinematics of the the BLR is
the ``reverberation mapping'' \citep{BM82, Ga88, Pe93, Pe04}, a method
that is based on multiple spectroscopic observation and the variability of spectral properties. 
In particular, the BEL's flux varies
with respect to the ionizing continuum flux with a certain time delay
(of the order of days to weeks), due to the light-travel time
from the source of the continuum photon to the source of the BEL photon, that is the BLR. 
Therefore, by applying the cross-correlation function (CCF) between the continuum
and BEL light curves an estimate of the time-lag $\tau$ between the two
signals can be obtained, which is basically the photometric BLR radius
$R_{\rm BLR} = c \tau$, where c is the speed of light. This is
a powerful tool to observe the unobservable, as one can estimate the size of the BLR 
and consequently the mass of the supermassive black hole, since the BLR gas is assumed to 
be virialized, and the method has been exploited in
many papers \citep[see e.g.][etc.]{Ka00, Pe04, Be09, Do12, Ko14}.
The reverberation-mapped AGN are of particular importance because they anchor the 
scaling relationships \citep[BLR radius-luminosity relationship, see e.g.][]{Ka00, Be06}
that allow the estimate of the BLR radius, and the supermassive black hole mass, from a single-epoch spectrum.  
 
However, this method is based on several assumptions, which makes it still
pretty uncertain, especially for the supermassive black hole mass estimate.
It is assumed that the BLR gas is photoionized by the 
continuum coming from the central source, and that is gravitationally bound to the supermassive
black hole so that the virial theorem can be used \citep{Ga88}.
Moreover, it strongly depends on the geometry of the BLR, and the kinematics and structure of
the BLR is not ubiquitously  described \citep[see e.g.][and reference therein]{Su00, Ga09}. 
The BLR  is probably very complex, often with evidence for 
multiple components \citep[e.g.][]{Su00, Il06, Po04, Bo09}, and can have systematic motions such as infalls,
outflows, circular motions, which should be detected in the BEL profiles. If the line profile is varying,
the change of the line shape should be according to the change in the BLR geometry and kinematics.
In principle, the reverberation assumption can be tested, e.g. 
one test is if the BEL fluxes correlate with the continuum flux. Therefore, detailed studies
of different broad line profiles is required, and this can be done with 
long-term (of the order of decade and longer) monitoring of type 1 AGN with 
different BEL properties.

In this paper we give a { comparative} overview of our investigation based
on the international long-term optical monitoring campaign,
coordinated by the Special Astrophysical Observatory of the Russian Academy of Science (SAO RAS).
Within this campaign, several type 1 AGN have been monitored for decades, out of which data were published
for 5 objects \citep{Sh01, Sh04, Sh08, Sh10a, Sh12, Sh13}. The objects have been
selected so that representatives of type 1 AGN with different BEL
profiles are in the sample: Arp 102B and 3C 390.3 have BELs with double-peaked line profiles, 
Ark 564 is a Narrow-Line Seyfert 1 (NLSy1), and NGC 4151 and NGC 5548 are
well-studied Seyfert 1.5 with strongly variable BEL profiles, with bumps and 
asymmetries present in the line profiles (see Figure 1). 
{ All 5 AGN have been discussed individually in referenced papers, but here
we aim to give for a first time a comparative analysis of their variabilty properties, 
that is to: i) summarize already obtained most important results of their BEL variability, ii) give new comparative analysis of the BEL properties, iii) discuss their complex BLR physics and geometry, all in the framework of the estimations of the mass of the supermassive black hole in AGN, and suggestions for future exploitation of our large data sets.} Our data sets are observed, reduced and analyzed using the same procedures, thus the total { analyzed} data form an uniform set of data.

The paper is organized as follows, in Section 2 we briefly report on observation
and data reduction, in Section 3 we present the main result, which are discussed in 
Section 4, and in Section 5 we outline our conclusions and future investigations.

\begin{figure}
\includegraphics[width=8.5cm]{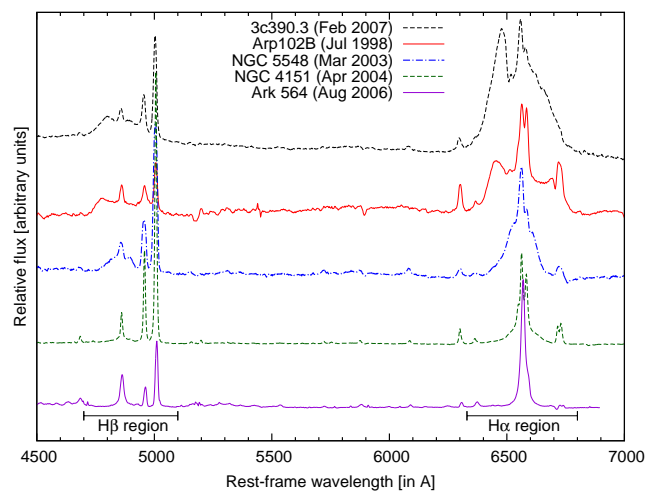}
\caption{Comparison of optical spectra for all 5 objects {(name and month of the observation
denoted)}, given in arbitrary flux unites and shifted for better display.
Rest-frame wavelengths are displayed on the X-axis, { and the spectral region of H$\beta$ and
H$\alpha$ line are marked}.} \label{fig1}
\end{figure}

\begin{figure}
\centering
\includegraphics[width=9cm]{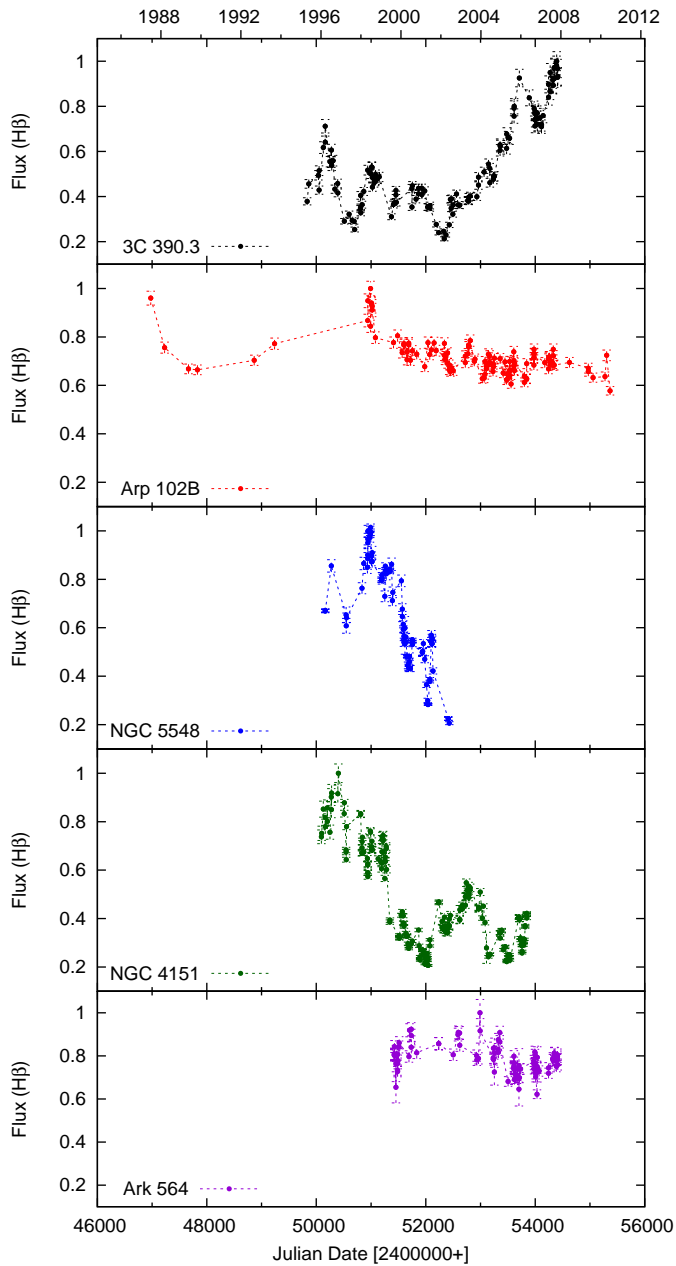}
\caption{H$\beta$ line light curves for all 5 objects (name denoted in bottom left).
The line flux { and corresponding error-bars} are normalized to the maximal flux for 
better comparison { and X-axis shows time in units of modified Julian Date.}} \label{fig2}
\end{figure}

\section{OBSERVATIONS AND DATA SETS}

With the international long-term optical monitoring campaign, coordinated by SAO RAS
several type 1 AGN have been monitored for decades, out of which data were published
for 5 objects in 2001--2013 (see Table 1 for details).
The high quality spectra were taken with 6 different telescopes: (i) 6-m and 1-m telescopes of SAO RAS; (ii)  2.1-m telescope of Guillermo Haro Astrophysical Observatory, Mexico; (iii) 2.1-m telescope of the Observatorio Astronomico Nacional at San Pedro Martir, Mexico, and (iv) the 3.5-m and 2.2-m telescopes of Calar Alto Observatory, Spain. The data acquisition, reduction and different calibration (e.g. corrections for different position angle, seeing and aperture
effects), and flux measurements were described in details in Shapovalova et al. (2001, 2004, 2008, 2010a, 2012, 2013). Table 1 gives a summary of the data (object name, monitored period, redshift and type, line properties, and references).

For all objects we measured different line parameters, { such as total line fluxes, 
broad line fluxes\footnote{To study the broad components showing the main BLR
characteristics, the narrow lines and the
forbidden lines were subtracted either by constructing observational templates or by
multi-Gaussian fittings.}, line widths, and line-flux time-lags, that will be investigated here 
(see references listed in Table 1 for details on other line parameters)}.
Some of the measured H$\beta$ and H$\alpha$ line parameters are given in Table 1: 
the full-width half-maximum (FWHM) of the mean spectrum, the photometric radius of the BLR 
$R_{\rm BLR} = c \tau_{\rm BLR}$, { where $\tau_{\rm BLR}$ is the line-flux time-lag} obtained from the CCF analysis, the maximal-to-minimal flux ratio $F_{\rm max} / F_{\rm min}$, the variability parameter $F_{\rm var}$, and the continuum luminosity at 5100 \AA. The variability parameter $F_{\rm var}$ is used to estimate the amount of the line-variability and is calculated according to \cite{OB98}. For the {estimates of time-lags from the} CCF analysis a Z-transformed Discrete Correlation Function (ZDCF) was used \citep{Al97}. Luminosities are calculated using the mean continuum flux at 5100 \AA \, and the online calculator for luminosity distance
\citep{Wr06}, for which we adopted the cosmological parameters $\Omega_{\rm } = 0.286$, $\Omega_{\Lambda} = 0.714$, and $\Omega_k = 0$, and a Hubble constant, $H_{\rm 0} = 69.6$ km/s Mpc$^{-1}$. 
All { flux measurements} from this campaign are publicly available in publications listed in Table 1 and in the corresponding VizieR Online Data Catalog.

The same procedures were performed for the observations, data reduction, measurements and analysis, so that the resulting data sets of these 5 objects basically are uniform. This makes them valid for further analysis and comparison.

\begin{figure}
\centering
\includegraphics[width=8.8cm]{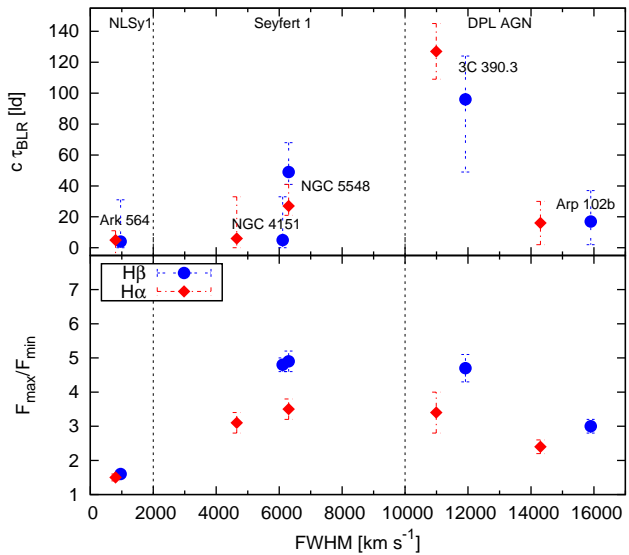}
\caption{Photometric radius of the BLR c$\tau_{\rm BLR}$ (upper) and 
the variability indicator, the ratio of maximal-to-minimal flux 
$F_{\rm max} / F_{\rm min}$ (bottom) versus the FWHM of H$\beta$ (circles)
and H$\alpha$ (diamonds) lines for all 5 objects.
The vertical dashed lines correspond to the line FWHM of
$2000$ and $10000\, {\rm km/s}$.} \label{fig3}
\end{figure}

\section{RESULTS}

Our sample consists of 5 type 1 AGN with different BEL
profiles (Figure 1): Arp 102B and 3C 390.3 have BELs with double-peaked line profiles, 
Ark 564 is a NLSy1, and NGC 4151 and NGC 5548 are
strongly variable Seyfert 1.5. The results of our long-term monitoring campaign
are summarized in Table 1, and Figure 1 and 2, that show the comparison of 
optical spectra for all 5 objects, given in
arbitrary flux unites and shifted for better display (Figure 1), and 
the H$\beta$ line flux light curves for all 5 objects (Figure 2), where
the line flux { and corresponding errors are} normalized to the maximal flux for better
comparison.

Similar as in \cite{Ko06}, in Figure 3 we plot the 
ratio of the maximal-to-minimal line flux against the FWHM of H$\beta$ (circles) and H$\alpha$ line (diamonds). 
{ We divided the $F_{\rm max} / F_{\rm min}$ vs. FWHM plane according to the line FWHM in three groups of objects: 1) NLSy1 having FWHM { $<$2000 km/s \citep{Os87}}, 2) AGN with broad emission lines (Seyfert 1) with 2000 $<$ FWHM $<$ 10000 km/s), and 3) double-peaked line AGN (DPL AGN) with FWHM $>$ 10000 km/s.} Figure 3 gives an indication that there are different types of objects (with different geometry and type of variability) that should be considered simultaneously. 

In order to test the photoionization
assumption, we plot in Figure 4 the correlation between the H$\beta$ line and continuum flux for  all objects (name indicated on each plot). The correlation coefficient and the corresponding P-value are listed in Table 2. The two objects, Arp 102B and Ark 564 have very low correlation coefficients (0.37 and 0.59, respectively), indicating a weak response of the line to the continuum variability.


\begin{table}
\caption{Line and continuum flux correlations for all 5 objects.
The correlation coefficient and the corresponding $P_{\rm null}$ value (in brackets) are given for each pair of data.}
\label{tab2}
\vskip 2mm
\centering
\begin{tabular}{ccc}
\hline
  object   &  H$\beta$ vs F$_{\rm cnt}$ & H$\alpha$ vs F$_{\rm cnt}$ \\
\hline
\hline          
NGC 5548  & 0.90 (0.1E-28) & 0.85 (0.2E-13)   \\
NGC 4151  & 0.94 (0.0) & 0.88 (0.0)   \\
3C390.3   & 0.91 (1.1E-19) & 0.82 (1.1E-12)  \\
Ark 564   & 0.59 (0.7E-9) & 0.71 (0.8E-8) \\
Arp 102B  & 0.37 (0.5E-4) & 0.25 (0.02)  \\
\hline
\end{tabular}
\end{table}

{ In principle, the geometry of the BLR should be seen in the line shape, e.g. in the most obvious case if the rotational motion is present in the BLR, then the line profile should have two peaks, the red corresponding to the gas moving away, and the blue one moving towards the observer. Moreover if there is a change in the BLR geometry and kinematics, the line shape should vary accordingly. In order to show how the line profiles vary in two DPL objects, we plot the mean and rms profile for the H$\beta$ line of 3C 390.3 (Figure 5, upper panel) and Arp 102B (Figure 5, bottom panel). From the shape and intensity of the rms profiles, it is obvious that in case of 3C 390.3 the line profiles vary significantly, while in Arp 102B the line profiles remain almost unchanged.}

Finally, in Figure 6, the radius of the BLR for H$\beta$ line versus the continuum luminosity
at 5100 \AA, is given. The solid line represents the scaling relation from \cite{Be06}, 
$\log R_{\rm BLR} = K + \alpha \log [\lambda L_\lambda(5100)/10^{44}]$, where $\alpha$ is the slope of the BLR radius-luminosity relationship and $K$ is the scaling factor, and { in Figure 6 $\alpha=0.533$ and $K=1.527$ from \cite{Be13} are plotted}.
The dashed line is a simple linear fit of the above equation (not considering the error-bars) through all objects except from Ark 564, obtaining the fitting parameters of $\alpha=0.887$ and $K=2.027$.

\begin{figure}
\centering
\includegraphics[width=8.8cm]{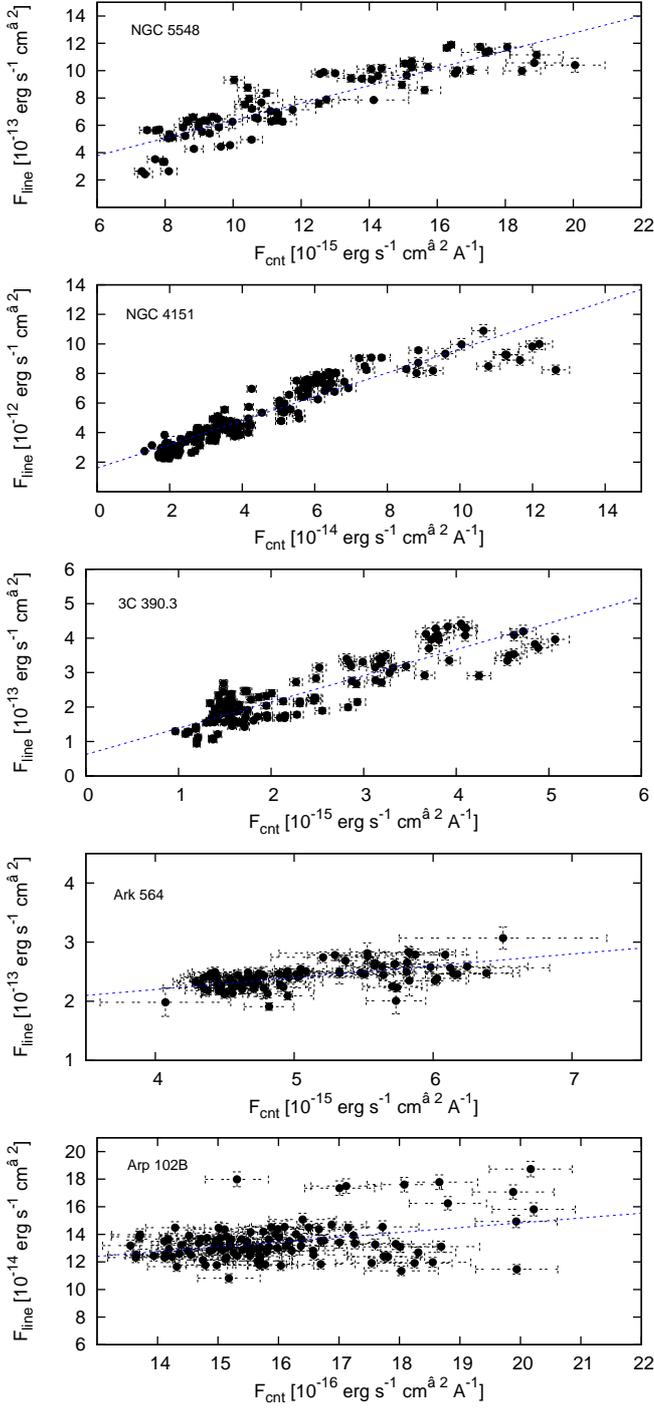}
\caption{Correlation between the line and continuum flux for the H$\beta$ line for all 
5 objects (name indicated on
each plot). The dashed line represents the linear best fit through all data.} \label{fig4}
\end{figure}

\begin{figure}
\centering
\includegraphics[width=8.8cm]{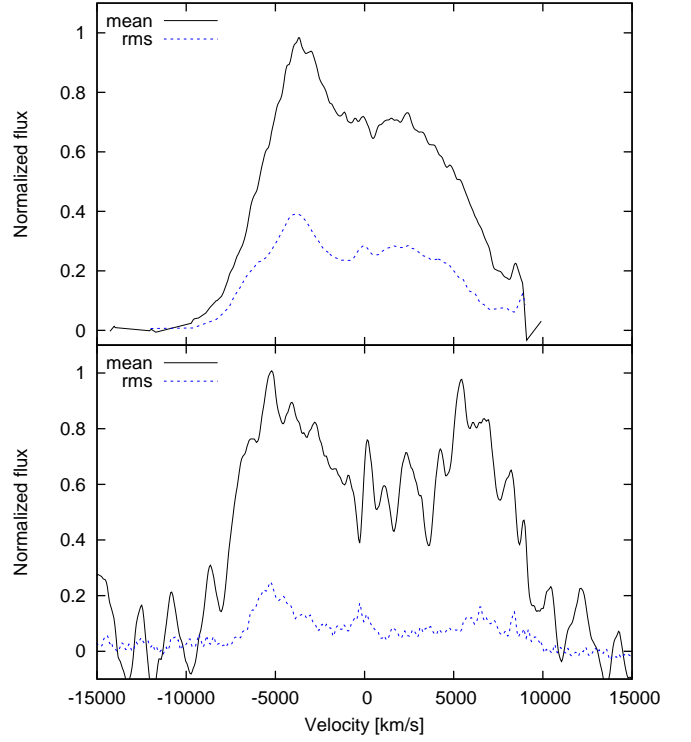}
\caption{Mean (solid line) and rms (dashed line) line profiles for the H$\beta$ line of 3C 390.3 (upper panel) and Arp 102B (bottom panel).} \label{fig5}
\end{figure}

\begin{figure}
\centering
\includegraphics[width=8.8cm]{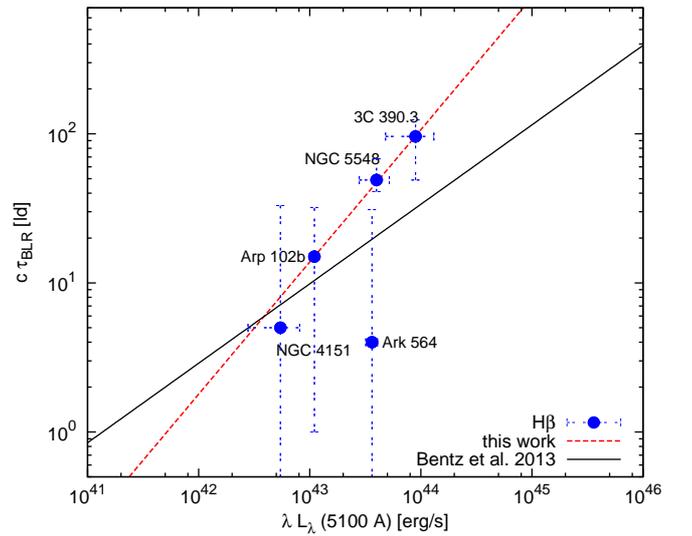}
\caption{The photometric radius of the BLR versus the continuum luminosity
at 5100 \AA, for H$\beta$ line. The solid line
represents the BLR radius-luminosity scaling relation from \cite{Be13} 
where $R_{\rm BLR}\sim L^{0.533}$, while the dashed line is our fit through all objects
except Ark 564, where $R_{\rm BLR}\sim L^{0.887}$.} \label{fig6}
\end{figure}

\section{DISCUSSION}

We have analyzed the variability of the optical spectra for five type 1 AGN with different properties of the broad emission lines. The data sets were processed in a uniform way, which makes them valid for further analysis and comparison.

During the long-term monitoring, that is by rule longer than a decade, the line and continuum flux for all objects are varying.
Most of the objects are strongly varying, i.e. NGC 5548, NGC 4151, and 3C 390.3, that is best seen from the normalized light curves (Figure 2), but also from the ratio of the maximal-to-minimal flux (Figure 3, Table 1) and the variability parameter F$_{\rm var}$  that is $\sim$40\% (Table 1). { This is supported with the correlation analysis, as these strongly varying objects are showing significant correlation between the line and continuum fluxes (Figure 4, Table 2). }

{ On the other hand, the line fluxes of the Ark 564, that is a NLSy1, stay almost constant during the monitored period (only changing by $\sim$7-8\%), that was also noticed before for this object \citep{She01}. It is also interesting  that in case of Arp 102B, that is the other extreme object with the largest line widths (FWHM $\sim$ 15000 km/s), the variability is also low ($\sim$20\%). 
The lack of variability in both of these cases is supported by the weak correlation
between the continuum and line fluxes (Figure 4, Table 2), e.g. for H$\beta$ of Arp 102B 
it is $r=0.37$ ($P_{\rm null}$=0.5E-04), and even worse for H$\alpha$ of the same object 
($r=0.25$, $P_{\rm null}$=0.22). The lack of variability can be due to poor sampling of the data,
e.g. in case of Arp 102B the mean sampling is 40 days \citep{Kov14}. However, the lack of correlation between the line and continuum flux also points to additional sources of ionization in the BLR apart from the central AGN continuum source. 
We should note here that our result for observed low variability  in 
NLS1 Ark 564 is in agreement with a comparative study of NLS1 and broad line Seyfert 1 galaxies (BLS1) 
variability given by \cite{ai13}. They found that the majority of NLS1-type AGN show variability on timescales from several days to a few years, but with the variabilty amplitudes
smaller than in the case of BLS1-type AGN. Moreover, 
there are some results in particular study cases that indicate
peculiar BLR in NLS1 AGN, as e.g. in NLS1 Mrk 493 \citep[see][]{Po09} where the broad lines 
have not changed in a period of couple of years (between
two observations), and it is suspected that the weak broad H$\alpha$ and 
H$\beta$  may come partly from the unresolved central BLR, but also may be partly produced by 
the violent starburst in the circumnuclear ring (clearly resolved on HST image). 
It seems that NLS1-type AGN have some specific physical 
properties that have to be considered in monitoring campaigns.}

{ Another peculiarity to note is the fact that the line flux appears to flatten at high continuum flux, that is especially seen in the case of NGC 4151 for which the line flux saturates above the continuum flux of $7\times 10^{−14} \rm erg \ cm^{-2} s^{-1} \AA^{-1}$ \citep[see thorough analysis and discussion in][]{Sh08}. Here this effect can be also noticed in the case of NGC 5548 and only weakly in 3C 390.3. This implys that either there is more ionizing to optical flux than expected for a typical AGN spectrum,
or that, again, lines are not produced purely by the photoionization from the central continuum source.}

The photometric radius of the BLR c$\tau_{\rm BLR}$ increases with the width of the BEL's profiles (Figure 3), except for the case of Arp 102B. This is in accordance with the general picture of the central engine, as it appears that in objects with lower accretion rate ($L/L_{\rm edd}$), those with larger line widths \citep{Ma14} the BLR is larger. Arp 102B though remains a mystery, as here some other geometry, apart from the obvious accretion disk geometry { that is usually} used to explain the double-peaked line profiles, may be considered \citep[see][]{Po14}. In favor to this is the fact that during more than two decades, the line profiles of this object have not changed significantly (Figure 5, bottom panel). This is not expected in case of the relatively compact accretion disk ($\sim$1000 gravitational radii) that was used to modeled these very broad double-peaked line profiles \citep[see models in][]{Ge07, Po14}. Additionally, there is an indication that an outflow in the BLR of Arp 102B is detected as shown in Figure 15 of \cite{Po14}. 

Finally, the radius-luminosity relationship plotted for these five objects shows that a linear trend is present (Figure 6). Obviously the NLSy1 Ark 564 is clearly an outliner. This is expected due to the low flux variability  of this object ($\sim$7\% for H$\beta$ line) and poor significance of the obtained BLR radius. Therefore, a simple fit was done through all objects except Ark 564, and we obtained $\alpha=0.887$ (Figure 6). This result is different from Bentz et al. (2013), who obtained $\alpha=0.553$, that is also plotted as solid line in Figure 6.  This illustrates that { considering only this very} low-luminosity end 
{ ($\lambda$L$_\lambda$(5100)$\sim 10^{43-44}$erg/s)} of the empirical radius-luminosity scaling relation is very sensitive { to the measured continuum luminosity and BLR radius. One reason could be the contribution of the host-galaxy continuum flux, that was not considered here, apart from subtracting the extended source correction factor $G(g)$, that is an aperture-dependent correction
factor standardly used in our campaign to account for the host-galaxy contribution. Moreover, the sample is far from being statistically significant. However, this still illustrates that new results of the monitoring campaigns may help in obtaining more accurate results and all AGN properties should be taken into account
(e.g. level of line variability or line production mechanism), as well as obtaining direct distances need for calculations of the luminosity as discussed in \cite{Be13}. }

It is obvious that for some objects, the BLR geometry is changing during decades of observation, e.g. in case of 3C 390.3 the rms profile of H$\beta$ shows strong variability especially in the blue peak \citep{Po11}. One possible scenario to model this variability is with the orbiting bright spots in accretion disk \citep{Jo10}. Other proposed scenario, e.g. for NGC 4151 that also has strongly variable BEL's profiles \citep{Sh10b} is that this well-studied object is hosting a binary supermassive black hole with two BLR orbiting around them \citep{Bo12}. { \cite{Bo12} found by analyzing the 20-year long NGC 4151 H$\alpha$ light curve and radial velocity curves of the line components the evidence for a sub-parsec scale supermassive binary black hole system with an orbital period of $\sim$5800 days.}


\section{CONCLUSIONS}

The { comparative study of} the main results of SAO RAS long-term
monitoring campaign of five type 1 AGN with BELs having different properties 
are given. The variations of the BELs parameters of
the H$\beta$ and H$\alpha$ line are investigated with the aim to constrain the
properties of the BLR and test the reverberation mapping as a method to estimate the size of the
BLR, and consequently the mass of the supermassive black hole in the center of an AGN.

The main conclusions are:
\begin{itemize}
\item[(i)] the five selected objects are varying during the monitored period, where some
are strongly variable ($\sim$40\%) in the spectral line fluxes (e.g. NGC 4151 and 3C 390.3);
\item[(ii)] { the BLR is believed to be dominantly photoionized by the central 
continuum source in the majority of AGN, but in some objects (in this analysis 3 out of 5, i.e. Arp 102B, Ark 564, and NGC 4151), other heating mechanisms such as a shock-ionization due to outflows, may contribute to the formation of the BEL};
\item[(iii)] the empirical radius-luminosity scaling relation is very sensitive, especially in this
low-luminosity end. It depends on the long-term reverberation monitoring results, so that these investigations continue to be important even though spectroscopy is a very telescope and time-consuming observational technique;
\item[(iv)] during decade(s) of monitored period, the size and
geometry of the BLR can change, and this should be considered when BELs are used  to
estimate the size of the BLR and the mass of the supermassive black hole.
\end{itemize}


The monitoring results presented here and in several other papers pay attention to the importance of the
spectroscopic monitoring campaigns in the investigation of the BLR properties and mass estimates of the
supermassive black holes in AGN.

\section*{Acknowledgments}
This work was supported by the Ministry of Education, Science and Technological Development
of Republic of Serbia through the project Astrophysical Spectroscopy of Extragalactic Objects (176001), INTAS (grant N96-0328), RFBR (grants
N97-02-17625 N00-02-16272, N03-02-17123, 06-02-16843, N09-02-01136,
12-02-00857a, 12-02-01237a, N15-02-02101), CONACYT research grants 39560, 54480,
and 151494. D.I. has been awarded L'Or{\'e}al-UNESCO "For Women in Science" National Fellowship for 2014.
{ We would like to thank the anonymous referees for very useful comments.}



\end{document}